\documentclass{article}

\usepackage[preprint]{neurips_2020}

\usepackage[utf8]{inputenc} 
\usepackage[T1]{fontenc}    
\usepackage{hyperref}       
\usepackage{url}            
\usepackage{booktabs}       
\usepackage{amsfonts}       
\usepackage{nicefrac}       
\usepackage{microtype}      
 
\usepackage{float}
\usepackage{booktabs}

\usepackage{textcomp}
\usepackage{gensymb}
\usepackage{nopageno}
\usepackage{times}
\usepackage{enumitem}
\usepackage[font=it]{caption}
\usepackage{doi}
\usepackage[usenames,dvipsnames]{xcolor}
\usepackage{listings}
\usepackage{array}
\usepackage{placeins}
\usepackage{multicol,titlesec,tabto, multirow, makecell}
\usepackage{longtable}

\usepackage{siunitx}
\usepackage{amsmath}
\usepackage{amsfonts}
\usepackage{amssymb}
\usepackage{mathtools}
\usepackage{eurosym, units}
\usepackage{setspace}
\usepackage{fancyvrb}
\usepackage{psfrag}

\newlength\figureheight
\newlength\figurewidth
\usepackage{tikz,pgfplots}
\usepackage{epsfig}
\usepackage{epstopdf}
\usepackage{graphicx}
\usepackage{caption}
\usepackage{subcaption}
\usepackage{color} 
\usepackage{siunitx}
\usepackage{cleveref}
\usepackage{natbib}

\title{LSTM-based Space Occupancy Prediction towards Efficient Building Energy Management}

\author{
Juye Kim \\
Electrical and Computer Engineering\\ 
Carnegie Mellon University\\
Pittsburgh, PA 15213 \\
\texttt{juyek@andrew.cmu.edu}\\
}

\begin{document}

\maketitle

\begin{abstract}

Energy consumed in buildings takes significant portions of the total global energy usage. A large amount of building energy is used for heating, cooling, ventilation, and air-conditioning (HVAC). However, compared to its importance, building energy management systems nowadays are limited in controlling HVAC based on simple rule-based control (RBC) technologies. The ability to design systems that can efficiently manage HVAC can reduce energy usage and greenhouse gas emissions, and, all in all, it can help us to mitigate climate change. 
This paper proposes predictive time-series models of occupancy patterns using LSTM. Prediction signal for future room occupancy status on the next time span (e.g., next 30 minutes) can be directly used to operate HVAC. For example, based on the prediction and considering the time for cooling or heating, HVAC can be turned on before the room is being used (e.g., turn on 10 minutes earlier). Also, based on the next room empty prediction timing, HVAC can be turned off earlier, and it can help us increase the efficiency of HVAC while not decreasing comfort. 
We demonstrate our approach's capabilities using real-world energy data collected from multiple rooms of a university building. We show that  LSTM's room occupancy prediction based HVAC control could save energy usage by 50\% compared to conventional RBC based control.

\end{abstract}

\section{Introduction}

Energy consumed in buildings takes significant portions of the total energy usage in the United States, approximately 40 percent of total energy \citep{energy2010energy}. Among the energy consumed in buildings, large parts are consist of heating, cooling, ventilation, and air-conditioning (HVAC). Hence, the ability to design systems and algorithms that can efficiently control HVAC can reduce energy usage and greenhouse gas emissions, helping us mitigate climate change.

However, compared to its importance, building energy management systems nowadays are limited in their ability to control HVAC based on simple rule-based control (RBC) technologies \citep{mechri2010use,aghemo2013management,serale2018model}. 

These systems control HVAC based on simple predefined conditions such as routine schedule or temperature thresholds. These simple rule-based controls are perhaps not the best way to increase the energy efficiency of building HVAC. Typically, temperatures of the room are controlled by HVACs on predefined scheduling (e.g., run only weekdays, working hours) and certain temperature setpoints. However, this predefined scheduling may not be the optimal approach as the actual usage of the building can be different from the predefined schedule.

More advanced building energy control can be attained by introducing data-driven predictive control approaches \citep{Jain8264315,DRGONA2018199,BUNNING2020109792}. Modern HVAC systems collect enormous building energy and environmental data in real-time \citep{fowler2017energy}. With these abundant building energy data such as energy usage, temperature, and human behavior data (e.g., room occupancy), we can build predictive models that can predict future energy usage and future human behavior patterns. These models can be used to enhance efficiency and effectiveness of control of HVAC  \citep{aftab2017automatic,afram2014theory,kim2018personal}. 

This paper proposes occupancy-based predictive building energy control using Long Short-Term Memory (LSTM) neural networks \citep{hochreiter1997long}.
We train the LSTM using the history of energy and environment data of a real-world building. 
For the prediction target variable, we concentrate on predicting room occupancy. Room occupancy prediction is a significant factor in building efficient energy management systems. Especially, human presence in a space is of great concern in accommodating optimal room environment (e.g., temperature, humidity, lighting, etc.). For instance, studies by Peng et al. \citep{peng2018using} showed occupancy-prediction-based cooling control could save 7–52\% of the energy used in office buildings as compared to scheduled cooling operations. Further, Wang and Chen \citep{wang2017energya,wang2017energyb} used indoor positioning systems (IPS) to obtain accurate occupancy distribution information across multiple spaces and simulated energy-saving in building air-conditioning control systems, and it showed about 22\% energy could be saved through the accurate occupancy information.

We validate our approach based on real-world energy usage data collected from a campus building. We conduct experiments with eight months of data points on multiple rooms from a building on campus. We first demonstrate that the LSTM-based approach can provide accurate predictions for future room occupancy states in multiple future time windows. Then, we demonstrate the capability of room occupancy prediction of LSTM could save energy usage by 50\% when it is used to control HVAC compared to conventional RBC.

\section{Approach}

\subsection{Multivariate Building Energy Time-Series and Prediction Task}

Energy and environmental data collected from a building can be represented as multivariate event time-series. All variables measured at one point in time can be regarded as an event record in multivariate event time-series. Specifically, at one-time point $t \in 1 , \dotsc, T$, all measured variables can consist of a multivariate event record vector $x_t \in \mathbb{R}^{n_x}$ where $n_x$ denotes the number of variables measured. With all data points collected throughout a span of time horizon with length $T$ we obtain multivariate event time-series $X = ({x}_1, \dotsc, {x}_T)$. 
Given multivariate event time series $X$, we can define a prediction task that, given a history of observations $X$, a model $M$ generates a prediction for the value of the target variable at the next time step: $\hat{y}_{t+1}$. In this paper, we set room occupancy state as the target variable $y$, and all previously measured variables $X$ are used as input to the model. $X$ includes airflow, actual room temperature,  setpoint temperature, cooling and heating setpoints, damper position command, discharge temperature, hot water valve command, actual space temperature, and room occupancy.

\subsection{LSTM-based Room Occupancy Prediction and Temperature Control} 

LSTM represents the current input $x_t$ into a hidden state vector $h_t$ through linear projections and non-linear activation functions. At the time of prediction, LSTM projects $h_t$ into the output target space, and with non-linear activation function, it generates the final prediction for the next time-step $\hat{y}_{t+1}$. At each time step, LSTM updates hidden states as follows: 
\begin{align*}
f_{t} &= \sigma(W^{(f)} \cdot [h_{t-1}, x_{t}] + b^{(f)}) & i_{t} &= \sigma(W^{(i)} \cdot [h_{t-1}, x_{t}] + b^{(i)}) \\
o_{t} &= \sigma(W^{(o)} \cdot [h_{t-1}, x_{t}] + b^{(o)}) & \Tilde{C}_{t} & = \text{tanh}(W^{c} \cdot [h_{t-1}, x_{t}] + b^{(c)}) \\
{C}_{t} &= f_{t} \cdot C_{t-1} + i_{t} \cdot \Tilde{C}_{t} & h_{t} &= o_{t} \otimes \text{tanh}(C_{t}) \\
\end{align*}
$f_{t}$, $i_{t}$, and $o_{t}$ are forget, input and output gates and $\otimes$ denotes element-wise multiplication. $\sigma$ denotes logistic sigmoid function and tanh denotes tangent hyperbolic function.

Future room occupancy prediction is generated through a fully-connected layer $W^{(fc)}, b^{(fc)}$ with output activation function sigmoid:
\begin{equation}
  \label{eq:pred_event_orig}
  \hat{y}_{t+1} = \sigma(W^{(fc)} \cdot h_{t} + b^{(fc)}) 
\end{equation}

Based on the room occupancy prediction $\hat{y}_{t+1}$, we can control HVAC for next time span (e.g., next 30 minutes) using setpoint scheduling \citep{FADZLIHANIFF201394}.

\section{Experiments Setup}
 
\vspace{-0.5cm}
\begin{table}[H]
\begin{center}
\scriptsize
\begin{tabular}{lrrrrr} \toprule {} &   Airflow Actual &   Airflow Setpoint &   Cooling Setpoint &   Damper Position Command &   Discharge Temperature \\ \midrule 
mean  &              62.46 &                69.26 &                74.77 &                       46.47 &                     62.39 \\ std   &              30.28 &                14.32 &                 3.34 &                       22.64 &                      5.31 \\ min   &              -2.38 &                65.00 &                72.00 &                       19.95 &                     55.04 \\ 
max   &             187.35 &               125.00 &                80.00 &                      100.00 &                    112.95 \\ \midrule \end{tabular} 

\begin{tabular}{lrrrr}
\toprule
{} &   Heating Setpoint &   HW Valve Command &   Occupancy Status &   Space Temperature Actual \\ \midrule 
mean  &                67.38 &                 1.24 &                 0.61 &                        71.04 \\ std   &                 1.83 &                10.86 &                 0.48 &                         2.06 \\ min   &                65.00 &                 0.00 &                 0.00 &                        67.09 \\ 
max   &                70.00 &               100.00 &                 1.00 &                        95.76 \\ \bottomrule 
\end{tabular} 
\caption{Data Statistics from one of the rooms that used for our experiment}
\vspace{-0.75cm}
\label{table-1}
\end{center}
\end{table}

\subsection{Dataset}
We obtained HVAC usage data from a university campus's energy data repository. Specifically, we use HVAC data on five rooms in the B-level of an academic and research purpose building. The HVAC data includes airflow (actual and setpoints), cooling and heating setpoints, damper position command, discharge temperature, HW valve command, actual space temperature, and occupancy status with a timestamp. The data were collected every minute, and we use a time span from 7/9/2019 to 2/1/2020. 

\subsection{Data Preprocessing}
We split the data into a train set and test set such that we use the first 70 percent of data points as a train set and use the rest of them as a test set. We use the following columns as a feature set $X$: airflow actual, airflow setpoint, cooling setpoint, heating setpoint, damper position command, discharge temperature, HW valve command, space temperature actual, day of the week, hour, month, and an indicator variable that represents weekend or non-weekend. For the prediction target $y$, we set occupancy status (binary). As shown in \Cref{table-1}, each feature has different mean and variance, we apply data normalization to real-valued features: $(x - \text{mean}) / \text{std}$.

\subsection{Experiment Details}
We set the dimension of hidden states of LSTM = 256, number of epochs = 10, and learning rate = 0.001, and update gradient at every 100 time steps. For optimizer, we use Adam \citep{kingma2014adam}. Our implementation is based on PyTorch and uses NVIDIA GTX 1080 Ti GPU to train all models. 

\subsection{Evaluation Metrics}
To evaluate the quality of predictions, we use the average precision, area under the receiver operating characteristic curve (AUROC), and binary cross-entropy (BCE) of prediction $\hat{y}_t$ and actual target value $y_t$ across all time steps $1,\dotsc, T$ in the test set.

\section{Results}

\begin{figure}[H]
\begin{center}
\subfloat[Binary Cross Entropy\label{fig:a}]{\includegraphics[scale=0.36]{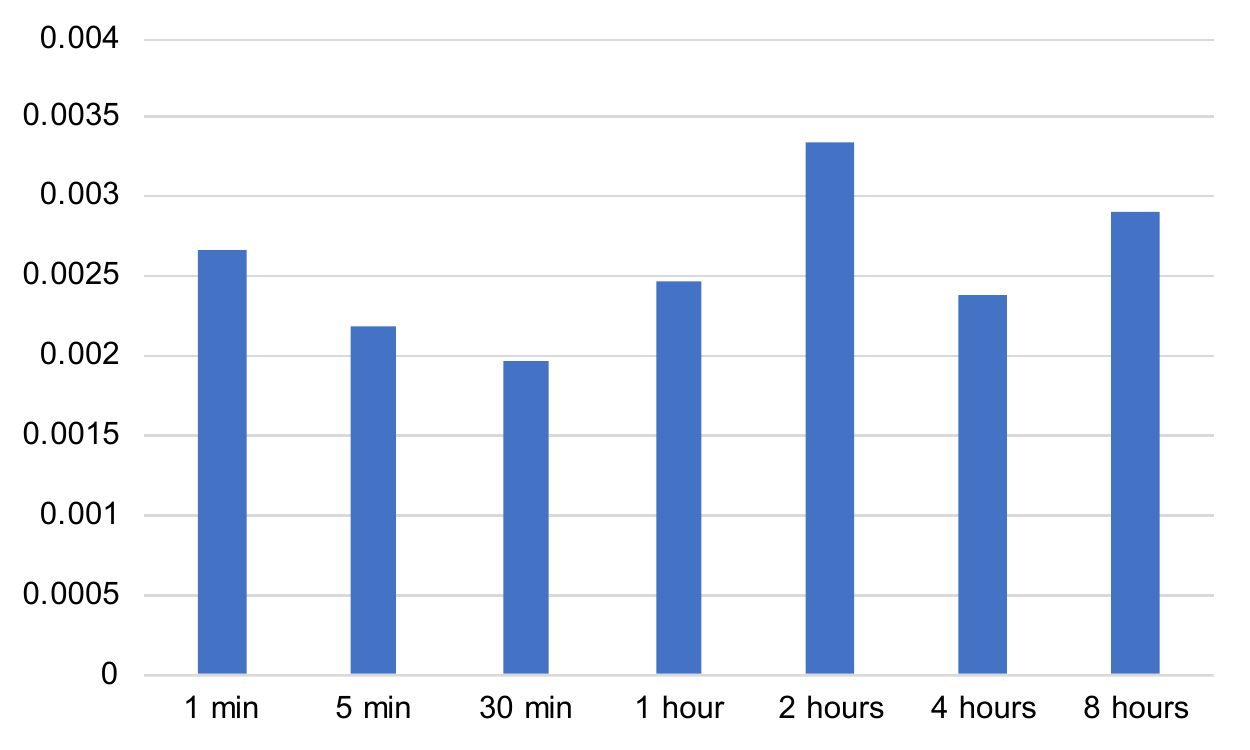}
}\hfill
\subfloat[AUROC\label{fig:b}]{\includegraphics[scale=0.36]{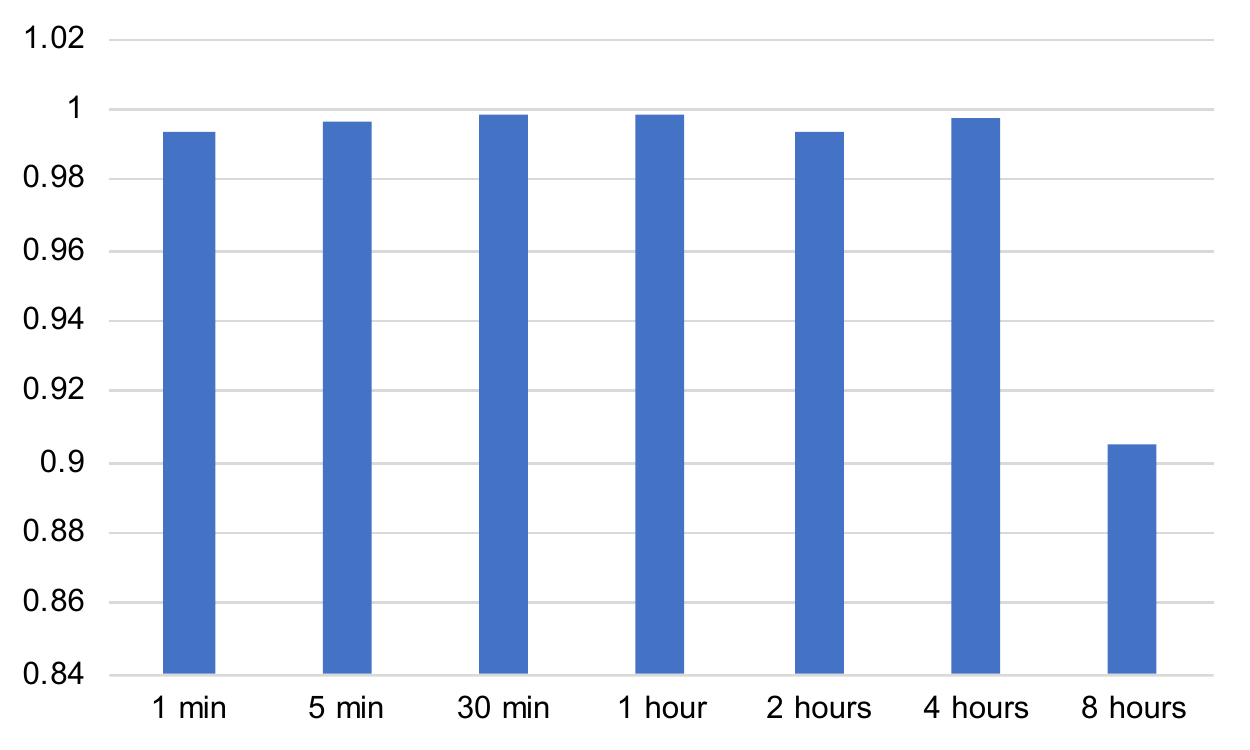}}\hfill
\subfloat[Average Precision\label{fig:c}]{\includegraphics[scale=0.36]{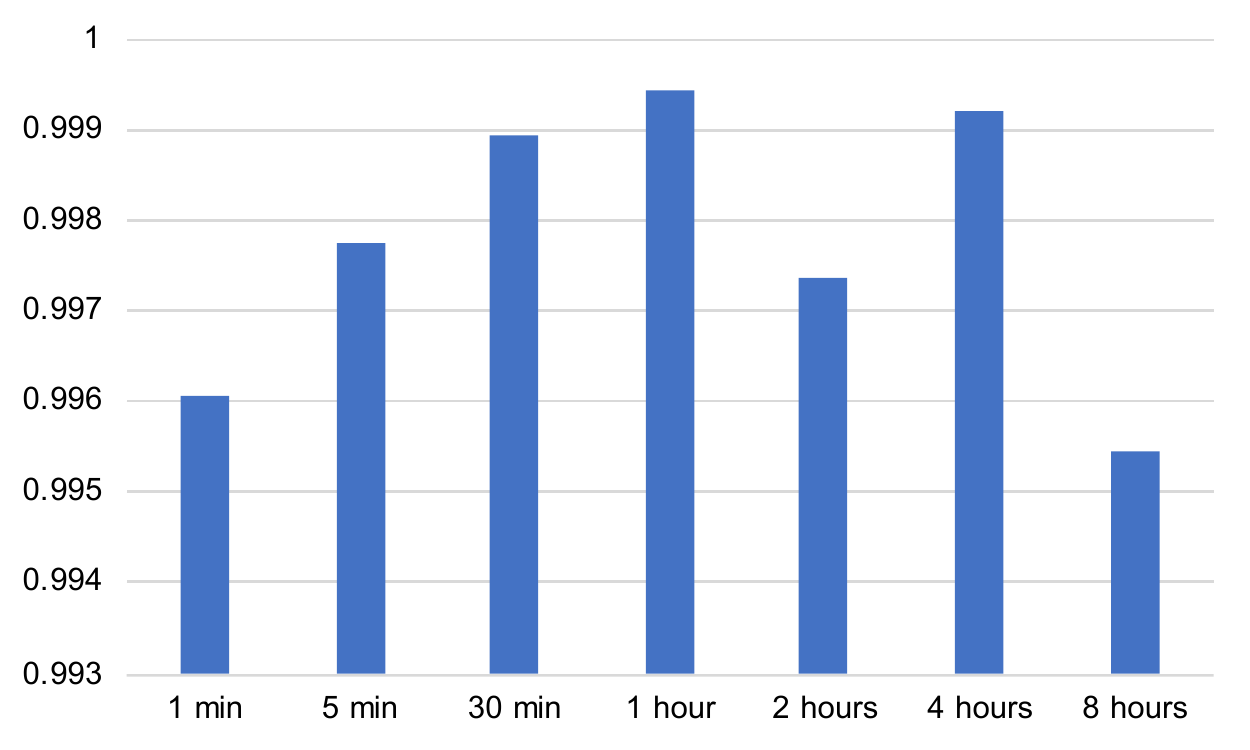}}\hfill

\caption{Performances of prediction on future time windows on varying sizes.}
\label{figure:result}
\end{center}
\end{figure}

\subsection{Task 1: Room Occupancy Prediction}
We evaluate our model on the task of predicting room occupancy of multiple next time windows. As the data is measured (sampled) every minute, given the history of all data, we evaluate the model's performance on predicting the next minute's room occupancy state. Especially, we aggregate future time points into varying sizes of time windows ($W$=1 minute, 5 minutes, 10 minutes, 1 hour, 2 hours, 4 hours, and 8 hours). For $W$ larger than 1 minute, we aggregate all future target variable states in the future time span ($W$) and represent it as a binary indicator variable ($1$ if $> 0$ else $0$). 

The prediction results are shown in \Cref{figure:result}. In terms of Binary Cross-Entropy (BCE) (\Cref{fig:a}) which the lower the better, we find predicting next 30 minutes shows the best performance and predicting 2 hours shows the worst performance. 
In terms of AUROC (\Cref{fig:b}), the models' performances across all prediction windows are similar except for 8 hours window. 
Meanwhile, in terms of average precision (\Cref{fig:c}), the best performance has shown at predicting the next 1 hour followed by predicting 30 minutes and 4 hours.
Looking at these three evaluation metrics, it seems using the prediction for room occupancy of the next 30 minutes or 1 hour would be a good choice. Also, considering a scenario that is based on the LSTM's room occupancy prediction, starting HVAC earlier or turning off HVAC earlier to increase the comfort of room environment and decrease unused energy, the size of future prediction window seems reasonable (e.g., 30 minutes).

\subsection{Task 2: Energy-efficient Occupancy-based Control}

\begin{table}[]
\begin{center}

\begin{tabular}{lrrr} 
    \toprule
    Space & Actual Energy Used by RBC & Saved Energy by LSTM &  Energy Savings \%
    \\ \midrule
RM-A & 43984.35        & 16514.52                                                & 37.55   \\
RM-B & 50959.11        & 30509.97                                                & 59.87   \\
RM-C & 49925.50        & 25850.20                                                & 51.78   \\
RM-D & 41466.86        & 20057.09                                                & 48.37   \\
RM-E & 34908.13        & 18200.54                                                & 52.14   \\ \midrule
Average & 44248.79        & 22226.46                                                & 49.94   \\ 
\bottomrule \end{tabular} 

\caption{Predicted energy savings (50\% in average) from LSTM's room-occupancy-prediction-based HVAC control, compared to simple rule-based control.}
\label{table-2}
\end{center}
\end{table}

We validate the effectiveness of the proposed occupancy-based control against the baseline of the simple rule-based control (RBC) method by comparing the energy usage of the two control methods. 
The energy use profiles of RBC are obtained from the measurement data of the task 1. Specifically, we compute the energy usage $E$ for HVAC based on convective heat equation given as a function of temperature set point $T_{sp}$, actual space temperature $T_{room}$, and airflow $\epsilon$: 
\begin{subequations}
\begin{align}
    \Delta T &= T_{\text{set point}} - T_{\text{actual space}} \\
    {E} &= C_{p} \cdot \epsilon \cdot \Delta T
\end{align}
\end{subequations}
where $C_{p}$ is specific heat capacity of air and we approximated it on a constant number at 300K (=74F) ($C_{p} = 1.005 kJ/kg.K)$. For temperature set point, we use heating set point from October to March and cooling set point for other months.
The energy use of the proposed occupancy-based scheduling control using LSTM is estimated based on the RBC's baseline energy use. Since the proposed control strategy will turn off the HVAC system during unoccupied periods, we estimate the amount of energy savings as RBC's energy use during predicted unoccupied periods by occupancy LSTM model. 
As shown in \Cref{table-2}, we can decrease energy usage by 50\% by controlling HVAC based on the proposed method, compared to conventional RBC based control.
Please note that these energy savings represent the theoretical performance bound of the proposed strategy.

\section{Conclusion}

This paper proposes building HVAC control based on room occupancy predictions based on LSTM. Apart from conventional rule-based controls, LSTM-based predictions can be used to directly control HVAC via setpoint scheduling by reflecting when a room will be occupied. 
We first demonstrated that using LSTM can provide a highly accurate prediction for next room occupancy states across multiple time windows as well as multiple rooms. By using the LSTM model as a predictive setpoint scheduler, we estimate that
presented occupancy-based control could decrease energy usage by up to 50\% compared to conventional RBC. 
Besides high projected energy savings, the primary added value of the proposed control method lies in its low cost and minimum expert requirements compared to competitive advanced control strategies.
The authors' future work includes the validation of the proposed control methodology in a real-world university campus building.

\section*{Acknowledgment}
The author thanks Ján Drgoňa (Pacific Northwest National Laboratory) for his significant contributions. He provided valuable guidance, suggestions, and insights to this work.


\bibliography{bibfile}
\end{document}